\documentclass[letterpaper,12pt]{article}
\usepackage{color,amssymb}

\usepackage{ifpdf}
\ifpdf
	\usepackage[pdftex]{graphicx}
	\usepackage[pdftex,unicode,implicit]{hyperref}

	\hypersetup{
  	pdftitle     = {Consistent Ho\v{r}ava gravity without extra modes and equivalent to general relativity at the linearized level},
  	pdfkeywords  = {},
  	pdfauthor    = {J. Bellor\'{\i}n, A. Restuccia and A. Sotomayor},
  	pdfcreator   = {pdf\LaTeXe\ with package \flqq hyperref\frqq},
  	pdfproducer  = {pdf\LaTeXe\ with package \flqq hyperref\frqq},
  	pdfpagemode  = UseNone,  
  	pdffitwindow = true,  
  	unicode      = true,
  	plainpages   = true,
  	colorlinks   = true,  
  	citecolor    = blue,  
  	urlcolor     = blue,
  	linkcolor    = blue
	}

\else
  \usepackage[dvips]{graphicx}
  \usepackage[unicode,implicit]{hyperref}

  \newcommand{\arXiv}[1]{{\tt arXiv:#1}}

\fi

\makeatletter
\@addtoreset{equation}{section}
\makeatother

\begin{document}

\begin{flushright}
{\small
Feb $21^{\rm th}$, $2013$}
\end{flushright}

\begin{center}

\vspace*{2cm}
{\bf \LARGE 
Consistent Ho\v{r}ava gravity without extra modes and equivalent to general relativity at the linearized level
} 
\vspace*{2cm}

{\sl\large Jorge Bellor\'{\i}n,}$^{a,}$\footnote{\tt jorgebellorin@usb.ve}
{\sl\large Alvaro Restuccia}$^{a,b,}$\footnote{\tt arestu@usb.ve}
{\sl\large and Adri\'an Sotomayor}$^{c,}$\footnote{\tt asotomayor@uantof.cl}
\vspace{3ex}

$^a${\it Departamento de F\'{\i}sica, Universidad Sim\'on Bol\'{\i}var, Valle de Sartenejas,\\ 
1080-A Caracas, Venezuela.} \\[1ex]
$^b${\it Department of Physics,}
$^c${\it Department of Mathematics, Universidad de Antofagasta, Antofagasta 1240000, Chile.}

\vspace*{2cm}
{\bf Abstract}
\begin{quotation}{\small
We consider a Ho\v{r}ava theory that has a consistent structure of constraints and propagates two physical degrees of freedom. The Lagrangian includes the terms of Blas, Pujol\`as, and Sibiryakov. The theory can be obtained from the general Horava's formulation by setting $\lambda = 1/3$. This value of $\lambda$ is protected in the quantum formulation of the theory by the presence of a constraint. The theory has two second-class constraints that are absent for other values of $\lambda$. They remove the extra scalar mode. There is no strong-coupling problem in this theory since there is no extra mode. We perform explicit computations on a model that put together a $z=1$ term and the IR effective action. We also show that the lowest-order perturbative version of the IR effective theory has a dynamics identical to the one of linearized general relativity. Therefore, this theory is smoothly recovered at the deepest IR without discontinuities in the physical degrees of freedom.
}\end{quotation}

\end{center}

\thispagestyle{empty}

\newpage
\section{Introduction}
The consistency of Ho\v{r}ava theory \cite{Horava:2009uw} is a subject that has been under intense study motivated by the search of a perturbatively renormalizable theory of quantum gravity. Numerous models that follow the original proposal of Ho\v{r}ava of adopting the foliation-preserving diffeomorphisms as gauge symmetry have been analyzed. An unavoidable question for all these models is whether general relativity (GR) can be consistently recovered at large distances, such that the Ho\v{r}ava theory can be regarded as the UV completion of GR. The originally proposed scheme for emergent GR inside Ho\v{r}ava theory \cite{Horava:2009uw} consists of obtaining the full action of GR (in the Arnowitt-Deser-Misner formulation) as the lowest-order effective IR action. This scheme requires the enhancing of the gauge symmetry group to the general space-time diffeomorphisms; that is, the original proposal is that not only the dynamics but also the symmetry of GR should be restored from Ho\v{r}ava theory in an approximate, effective way. To achieve this, the IR effective action should satisfy the condition that the coupling constant $\lambda$ multiplying the trace-kinetic term,
\begin{equation}
 \sqrt{g} N ( K_{ij} K^{ij} - \lambda K^2 ) \,,
\end{equation}
must approach the value $\lambda=1$ in order for this combination to match its fully covariant version.

An alternative to this scheme was noticed in Ref.~\cite{Bellorin:2010je}, where it was shown, by means of a Hamiltonian analysis, that the lowest-order truncation of the original, nonprojectable, Ho\v{r}ava theory, which is given by the Lagrangian
\begin{equation}
 \sqrt{g} N (K_{ij} K^{ij} - \lambda K^2 + R) \,, 
\end{equation}
is physically equivalent to a gauge-fixed version of general relativity (the gauge in which $K=0$, the so-called maximal slicing gauge). This means that both theories are dynamically identical, although their gauge symmetry groups are different. The crucial point for the result of Ref.~\cite{Bellorin:2010je} is that the condition $K=0$ emerges as one of the constraints of the theory, hence $\lambda$ becomes meaningless for the IR effective action. In particular, this result shows how it is possible to get GR without the requisite $\lambda \rightarrow 1$.

However, there is a central question about the discontinuities that might arise in recovering GR. Since Ho\v{r}ava theory has a reduced gauge symmetry group, generically it has an extra degree of freedom with respect to GR (this was already studied in Ref.~\cite{Horava:2009uw}). There is an abundant quantity of works devoted to the physics of this extra mode. We may mention that in Ref.~\cite{Charmousis:2009tc} it was signaled the problem of its strong coupling in the original theory; implying, instead of a discontinuity, the breaking down of the whole perturbative analysis. We want to stress that this result is based upon the assumption that $\lambda \rightarrow 1$ at the IR as a generic rule to get GR. In Ref. \cite{Blas:2009yd}, using a curvature-square model, it was shown that the extra mode is of odd nature, that is, propagates itself with a first-order time derivative (see also \cite{Li:2009bg}). This was confirmed in Ref.~\cite{Bellorin:2010te}, using again a curvature-square model, but showing also that the algebra of constraints closes and that the extra mode decouples at large distances smoothly in perturbative analysis\footnote{The results of Refs.~\cite{Bellorin:2010je,Bellorin:2010te} were corroborated in Ref.~\cite{Das:2011tx}.} \footnote{A perturbative analysis in a projectable model of the theory can be found in Ref.~\cite{Bogdanos:2009uj}.}.

With the aim of curing the oddness of the extra mode, it was noticed in Ref.~\cite{Blas:2009qj} that the nonprojectable Ho\v{r}ava action admits a large class of new terms (once the principle of detailed balance is discarded). Since the vector $a_i = \partial_i \ln N$ is covariant under the foliation-preserving diffeomorphisms, scalar combinations of it and the spatial metric are admissible into the Lagrangian. Adopting the logic of renormalizable gauge field theories, all the terms that are compatible with the gauge symmetry must be included in the Lagrangian. This leads us to consider the complete, nonprojectable, Ho\v{r}ava theory as the one containing the terms of Blas, Pujol\`as, and Sibiryakov \cite{Blas:2009qj}. Those authors found in \cite{Blas:2009qj} that the extra mode becomes even (propagates with a second-order time derivative) in the complete theory. However, the authors of Ref.~\cite{Papazoglou:2009fj} reported that the strong coupling problem persists in the complete theory, assuming again the condition $\lambda \rightarrow 1$ at the IR (see also \cite{Kimpton:2010xi}). It has been argued \cite{Blas:2009ck} that this problem can be avoided by requiring that the scale at which high-order operators become relevant is low enough.

On the side of the Hamiltonian analyses, in Refs.~\cite{Kluson:2010nf,Donnelly:2011df,Bellorin:2011ff} it was studied the closure of the algebra of constraints of the complete theory. In Refs.~\cite{Donnelly:2011df,Bellorin:2011ff} explicit computations were performed on the lowest-order IR effective action (second-order in derivatives). It was shown that the algebra of constraints closes: in particular well-behaved elliptic partial differential equations arise (see \cite{Bellorin:2012di}), and these analyses confirm that the theory has an extra even scalar mode, which is persistent at the level of the IR effective action.

It is important to realize that there is an underlying hypothesis behind these Hamiltonian analyses: it is assumed that the time derivatives $\dot{g}_{ij}$ can be completely solved in terms of the momenta $\pi^{ij}$ \cite{Bellorin:2011ff}. This is effectively the case when $\lambda$ satisfies $\lambda \neq 1/3$. However, when $\lambda = 1/3$ it is not possible to solve $\dot{g}_{ij}$ completely in terms of $\pi^{ij}$; instead, the primary constraint $g_{ij} \pi^{ij} = 0$ emerges. Since this constraint is not present in the case $\lambda \neq 1/3$, it is to be expected that the number of physical degrees of freedom is reduced when $\lambda = 1/3$. Therefore, it becomes of great interest to study the Hamiltonian formulation for the complete Ho\v{r}ava theory under the special value $\lambda = 1/3$, since this case might be an exception to the generic presence of the extra mode\footnote{The Hamiltonian formulation of the lowest-order truncation of the original, nonprojectable, Ho\v{r}ava theory at the value $\lambda =1/3$ was carried out in Refs.~\cite{Pons:2010ke,Bellorin:2010je}.}. Moreover, one may ask whether the dynamics of theory at $\lambda = 1/3$ is closer to the one of GR than for other values of $\lambda$. We emphasize that the $g_{ij} \pi^{ij} = 0$ constraint protects the value $\lambda =1/3$, since any other value for $\lambda$ would imply the violation of this constraint and the quantum formulation must be done on the constrained submanifold.

With these goals in mind, in this paper we perform the Hamiltonian analysis to the complete, nonprojectable Ho\v{r}ava theory fixing the special value $\lambda = 1/3$. Explicit computations on the complete Ho\v{r}ava theory are very difficult since the Lagrangian has a big number of higher-order terms. To overcome this difficulty, our first strategy consists of dealing with a general potential and making computations in an implicit form. This will allow us to arrive at conclusive results on the closure of the algebra of constraints and the number of physical degrees of freedom. Then we move to a specific model which has a $z=3$ term and the most general $z=1$ terms. The $z=3$ term is the square-Cotton tensor term and can be obtained by the detailed balance principle \cite{Horava:2009uw,Griffin:2011xs}. The $z=1$ terms give the relevant action for the large-distance dynamics, the IR effective action. Of course, this is justified by the assumption that all higher-order terms are suppressed at low energies. This model, which will allow us to perform explicit computations, can be regarded as a theory with soft breaking of conformal symmetry, because the square-Cotton term is conformally invariant whereas, the $z=1$ terms break the conformal symmetry. 

We give in advance our three main results: \emph{The Ho\v{r}ava theory we consider has a closed structure of constraints, propagates two physical degrees of freedom, and the linearized version of its IR effective action coincides with linearized GR}. Notice that these results imply that linearized GR is recovered at the lowest energies without discontinuities in the physical degrees of freedom. For the concrete model we analyze, we also found that \emph{in a sector of the space of parameters the energy of the model is nonnegative}\footnote{Positiveness theorems for the energy of nonprojectable Ho\v{r}ava theory in the case $\lambda \neq 1/3$ have been formulated in Refs.~\cite{Bellorin:2012di,Garfinkle:2011iw}.}.


\section{Hamiltonian analysis of the full theory}
We now start with the computations. Most of the steps are parallel to the ones of the case $\lambda\neq 1/3$. In order to have a self-contained study, we shall perform the whole analysis from the very beginning, making special emphasis on the results that depart from the $\lambda\neq 1/3$ case. 

The action of the complete, nonprojectable Ho\v{r}ava theory is written in terms of the Arnowitt-Deser-Misner (ADM) variables $g_{ij}$, $N$ and $N_i$ as
\begin{equation}
 S = \int dt d^3x \sqrt{g} N 
       \left( G^{ijkl} K_{ij} K_{kl} - \mathcal{V} \right),
\label{lagrangianaction}
\end{equation}
where
\begin{eqnarray}
K_{ij} & = & \frac{1}{2N} ( \dot{g}_{ij} - 2 \nabla_{(i} N_{j)} ) \,,
\\[1ex]
G^{ijkl} & = &
\frac{1}{2} \left( g^{ik} g^{jl} + g^{il} g^{jk} \right) 
- \lambda g^{ij} g^{kl}
\end{eqnarray}
and the potential $\mathcal{V}=\mathcal{V}(g_{ij},a_i,\ldots)$ is the most general combination of the spatial metric, its curvature tensor, the vector $a_i$ and covariant spatial derivatives of these objects that transforms as a scalar under spatial diffeomorphisms. To ensure power-counting renormalizability, the potential must include at least terms of order $z=3$, which means that they are of sixth order in spatial derivatives. The potential can also include a cosmological-constant term; we put it equal to zero, hence our simplest vacuum is the Minkowski space-time. The lowest-order terms, which yield the effective action for large distances, are
\begin{equation}
 \mathcal{V}^{(2)} = - R - \alpha a_i a^i \,.
\label{quadraticpotential}
\end{equation}
$\lambda$ and $\alpha$ are coupling constants of the theory.

By expanding the kinetic term we get $G^{ijkl} K_{ij} K_{kl} = K_{ij} K^{ij} - \lambda K^2$, where $K = g^{ij} K_{ij}$. If we were dealing with GR we were forced to put $\lambda = 1$, since only the combination $K_{ij} K^{ij} - K^2$ is covariant under transformations mixing time with space. However, under foliation-preserving diffeomorphisms, both $K_{ij} K^{ij}$ and $K$ are separately covariant, hence $\lambda$ is left undetermined by the gauge symmetry of the Ho\v{r}ava theory. From the general relation
\begin{equation}
 G^{ijkl} g_{kl} = (1 - 3\lambda) g^{ij} \,,
\end{equation}
we may see that the value $\lambda = 1/3$ is special since the metric becomes a null eigenvector of the four-index metric $G^{ijlk}$, 
\begin{equation}
G^{ijlk} g_{kl} = 0 \,.
\label{nullG}
\end{equation}
This implies that the metric $G^{ijkl}$ is not invertible for $\lambda = 1/3$.

Let us perform the Legendre transformation to the action (\ref{lagrangianaction}) in the case $\lambda = 1/3$. Since general spatial diffeomorphisms are part of the gauge symmetries of the theory, we know that the shift functions $N_i$ can be regarded as the Lagrange multipliers associated to the first-class constraint generating this kind of transformations. Hence the phase space is spanned by the conjugated pairs $(g_{ij}, \pi^{ij})$ and $(N,\phi)$. The action (\ref{lagrangianaction}) does not depend explicitly on the time derivative of $N$, hence we get the primary constraint
\begin{equation}
 \phi = 0 \,.
\end{equation}
The momentum conjugated to the spatial metric is given by
\begin{equation}
\frac{\pi^{ij}}{\sqrt{g}} = 
G^{ijkl} K_{kl} \,.
\label{momentum}
\end{equation}
By using (\ref{nullG}) and denoting $\pi \equiv g_{ij} \pi^{ij}$, we get the primary constraint
\begin{equation}
  \pi = 0 \,.
\label{picero}
\end{equation}
This constraint is absent in the complete, nonprojectable Ho\v{r}ava theory with $\lambda\neq 1/3$ \cite{Kluson:2010nf,Donnelly:2011df,Bellorin:2011ff}. $\pi$ is the generator of conformal transformations on $g_{ij}$ and $\pi^{ij}$. However, these are not part of the gauge symmetries of the theory, hence we may anticipate that $\pi = 0$ is a second-class constraint\footnote{For the special case of a potential containing only the square of the Cotton tensor the action acquires a conformal symmetry at $\lambda = 1/3$ \cite{Horava:2009uw}.}. A similar consideration applies for $\phi = 0$.

From (\ref{momentum}) it is straightforward to obtain the relations
\begin{eqnarray}
 G^{ijkl} K_{ij} K_{kl} & = &
 \frac{1}{g} \pi^{ij} \pi_{ij}  \,,
\\
\pi^{ij} \dot{g}_{ij} & = &
  \frac{2 N}{\sqrt{g}} \pi^{ij} \pi_{ij} + 2 \pi^{ij} \nabla_i N_j \,,
\end{eqnarray}
and using them we may build the Hamiltonian. We get, after an integration by parts that yields no boundary contributions,
\begin{equation}
 H = \int d^3x \left(
       \frac{N}{\sqrt{g}} \pi^{ij} \pi_{ij} + \sqrt{g} N \mathcal{V}
       + N_i \mathcal{H}^i + \sigma \phi + \mu \pi 
           \right) 
     + E_{\mbox{\tiny ADM}} \,,
\label{H}
\end{equation}
where $\mathcal{H}^i = 0$ is a primary constraint defined by
\begin{equation}
 \mathcal{H}^i \equiv - 2 \nabla_j \pi^{ij} + \phi \partial^i N \,.
\label{momentumconstraint}
\end{equation}
This is the so-called momentum constraint, which is the generator of spatial diffeomorphisms, hence it is a first-class constraint. The term proportional to $\phi$ in $\mathcal{H}^i$ ensures having the complete generator, since in the complete Ho\v{r}ava theory $N$ is part of the canonical variables. We may add this term since $\phi=0$ is a constraint of the theory. We have also added the rest of primary constraints to the Hamiltonian (\ref{H}), such that $N_i$, $\sigma$ and $\mu$ enter as Lagrange multipliers. Obviously, $\mu$ does not arise in the case $\lambda\neq 1/3$. The ADM energy,
\begin{equation}
  E_{\mbox{\tiny ADM}} = \oint d\Sigma_i ( \partial_j g_{ij} - \partial_i g_{jj} ) \,,
\end{equation}
is included \emph{a la} Regge and Teitelboim \cite{Regge:1974zd} in order to obtain the equations of motion from the most general variations of the Hamiltonian that are compatible with the boundary conditions.

Now we study the preservation in time of the primary constraints. Since $\mathcal{H}^i = 0$ is a first-class constraint, we concentrate ourselves in the preservation of $\phi = 0$ and $\pi = 0$. For the time evolution of $\phi$, we need to compute its Poisson bracket with the Hamiltonian (\ref{H}).  In computing the Poisson brackets we can omit the boundary terms of the Hamiltonian, since they do not give local contributions to these brackets. We just need to remember that we may discard any nonzero boundary contribution arising in the derivative of the bulk terms, since the boundary terms of the Hamiltonian account for them. Notice that arbitrary variations of the potential with respect to $N$ can be written in a closed form,
\begin{eqnarray}
\delta_N \int d^3x \sqrt{g} N \mathcal{V} &=& 
\int d^3x \sqrt{g} \left( 
   \mathcal{V} \delta N
   + N \frac{\partial \mathcal{V}}{\partial a_i} \delta a_i
   + N \frac{\partial \mathcal{V}}{\partial (\nabla_j a_i)} 
                                               \delta \nabla_j a_i
   + \cdots
   \right)
\\ 
&=& \int d^3x \sqrt{g} \left( 
   \mathcal{V} \delta N
   + N \sum\limits_{r=1} 
      \frac{\partial \mathcal{V}}{\partial ( \nabla_{i_r \cdots i_2} a_{i_1} )} 
         \nabla_{i_r \cdots i_2} \delta a_{i_1}
   \right) \,,
\label{varv}
\end{eqnarray}
where we are using the shorthand $\nabla_{ijk\cdots} \equiv \nabla_i \nabla_j \nabla_k \cdots$. After using $\delta a_i = \partial_i (\delta N / N)$ and integrating by parts (no term gives boundary contributions), we get
\begin{equation}
\delta_N \int d^3x \sqrt{g} N \mathcal{V} = 
\int d^3x \sqrt{g} \delta N \left[ 
     \mathcal{V} 
   + \frac{1}{N} \sum\limits_{r=1} (-1)^r 
      \nabla_{i_1 \cdots i_r} \left( N 
         \frac{\partial \mathcal{V}}{\partial ( \nabla_{i_r \cdots i_2} a_{i_1} )} 
          \right)
   \right] \,.
\end{equation}
Therefore, the condition $\{\phi,H\} = 0$ leads to the secondary constraint $\mathcal{H} = 0$, where
\begin{equation}
 \mathcal{H} \equiv
 \frac{1}{\sqrt{g}} \pi^{ij} \pi_{ij}  
 + \sqrt{g} \tilde{\mathcal{V}} 
\label{hamiltonianconstraintgeneral}
\end{equation}
and we have introduced the modified potential
\begin{equation}
\tilde{\mathcal{V}} \equiv
\mathcal{V} 
   + \frac{1}{N} \sum\limits_{r=1} (-1)^r 
      \nabla_{i_1 \cdots i_r} \left( N 
         \frac{\partial \mathcal{V}}{\partial ( \nabla_{i_r \cdots i_2} a_{i_1} )} 
          \right) \,.
\label{modifiedpotential}
\end{equation}
Following the standard nomenclature of GR, we call $\mathcal{H} = 0$ the Hamiltonian constraint.

The Hamiltonian given in (\ref{H}) can be written in terms of $\mathcal{H}$, the other constraints and boundary terms. To achieve this, we notice that an integration by parts and the behavior of the fields at infinity allow us to verify the identity
\begin{equation}
\int d^3x N \mathcal{H} =
\int d^3x \left(
       \frac{N}{\sqrt{g}} \pi^{ij} \pi_{ij} + \sqrt{g} N \mathcal{V} \right)
+ 2 \alpha \Phi_N \,,
\label{intnh}
\end{equation}
where $\Phi_N$ is the flux of $N$ at spatial infinity,
\begin{equation}
 \Phi_N \equiv \oint d\Sigma_i \partial_i N \,.
\end{equation}
To arrive at the integral (\ref{intnh}) we have used the fact that all the derivatives of the potential enter in $N\mathcal{H}$ inside total divergences; the only one that does not vanish when integrated at spatial infinity is the derivative of the quadratic term in $a_i$, which yields $2 \alpha \sqrt{g} N a^i =\mathcal{O}(r^{-2})$. Its surface integral at infinity can be further simplified such that it gives the flux of $N$. On the basis of (\ref{intnh}), we may write the Hamiltonian (\ref{H}) as a sum of constraints plus boundary terms,
\begin{equation}
 H = \int d^3x \left(
       N \mathcal{H} + N_i \mathcal{H}^i + \sigma \phi + \mu \pi 
           \right) 
     + E_{\mbox{\tiny ADM}} - 2 \alpha \Phi_N \,.
\label{hamiltonianfinal}
\end{equation}
From this expression, it is clear that the energy of the theory is given by
\begin{equation}
 E = E_{\mbox{\tiny ADM}} - 2 \alpha \Phi_N \,.
\label{e}
\end{equation}
From a mathematical point of view, the role of the boundary term proportional to $\Phi_N$ in the Hamiltonian (\ref{hamiltonianfinal}) is analogous to the one of the ADM energy \cite{Regge:1974zd}: it ensures the differentiability of the Hamiltonian under variations of $N$ that behave as $\delta N = \mathcal{O}(r^{-1})$ at infinity. In particular, we shall see below that there is a term proportional to $\nabla_i a^i$ in the modified potential $\tilde{\mathcal{V}}$. Its variation with respect to $N$ gives rise to a nonzero boundary term that cancels out with the variation of $-2\alpha\Phi_N$.

We now turn our attention to the preservation of $\pi=0$. We find that the condition $\{\pi,H\}=0$ leads to the constraint $\mathcal{C}=0$, where
\begin{equation}
\mathcal{C} \equiv
\frac{3N}{2\sqrt{g}} \pi^{ij} \pi_{ij} 
- \sqrt{g} \tilde{\mathcal{V}}\,' \,,
\label{cconstraint}
\end{equation}
and we have introduced the objects
\begin{equation}
\sqrt{g} \tilde{\mathcal{V}}\,'{}^{ij} \equiv \frac{\delta}{\delta g_{ij}} \int d^3y \sqrt{g} N \tilde{\mathcal{V}} \,,
\hspace{2em}
\tilde{\mathcal{V}}\,' \equiv g_{ij} \tilde{\mathcal{V}}\,'{}^{ij} \,.
\label{vprima}
\end{equation}
Again, this constraint is absent in the complete theory with $\lambda\neq 1/3$. Notice that there are some similarities in the structures of $\mathcal{H}$ and $\mathcal{C}$. Two immediate consequences we obtain from the vanishing of them are
\begin{equation} 
\tilde{\mathcal{V}} \leq 0 \,,
\hspace{2em}
\tilde{\mathcal{V}}\,' = - \frac{3}{2} N \tilde{\mathcal{V}} \,.
\label{inequalities}
\end{equation}

The next step is to impose the preservation in time of the secondary constraints $\mathcal{H}=0$ and $\mathcal{C}=0$. By computing Poisson brackets with the Hamiltonian, we obtain that the condition $\{\mathcal{H},H\}=0$ leads to the equation
\begin{equation}
\begin{array}{rcl}
{\displaystyle
\int d^3y
  \,\sigma\, \frac{\delta}{\delta N} \int d^3z \sqrt{g} \tilde{\mathcal{V}} \delta
+ \int d^3y \,\mu\, g_{ij} \frac{\delta} 
       {\delta g_{ij}} \int d^3z \sqrt{g} \tilde{\mathcal{V}} \delta
- \frac{3 \pi^{ij} \pi_{ij}}{2 \sqrt{g}} \mu }  
\hspace{2em} & & \\[1ex]
{\displaystyle
+ 2 \int d^3y \frac{N \pi_{ij}}{\sqrt{g}} 
     \frac{\delta }{\delta g_{ij}} \int d^3z \sqrt{g} \tilde{\mathcal{V}} \delta
- 2 \pi^{ij} \tilde{\mathcal{V}}\,'_{ij} }
& = & 0 \,.
\label{sigmaeq}
\end{array}
\end{equation}
The symbol $\delta$ is the Dirac delta centered at the point $x$ at which this equation is evaluated, $\delta = \delta^3(z-x)$ \footnote{Let us further explain the notation. For example, the first term of (\ref{sigmaeq}) should be read as
\[
 \int d^3y
  \,\sigma(y) \, \frac{\delta}{\delta N(y)} \int d^3z \sqrt{g(z)} \tilde{\mathcal{V}}(z) \delta^3(z-x)
\]
and similarly the other terms of (\ref{sigmaeq}) and (\ref{mueq}) that involve integrals and functional derivatives.}. Similarly, the condition $\{\mathcal{C},H\}=0$ leads  us to the equation
\begin{equation}
\begin{array}{rcl}
{\displaystyle
  \int d^3y \,\mu\, g_{ij} \frac{\delta}{\delta g_{ij}} 
     \int d^3z \sqrt{g} \tilde{\mathcal{V}}\,' \delta  
+ \int d^3y \,\sigma\, \frac{\delta}{\delta N} 
     \int d^3z \sqrt{g} \tilde{\mathcal{V}}\,' \delta
+ \frac{9 N \pi^{ij} \pi_{ij}}{4\sqrt{g}} \mu  
- \frac{3 \pi^{ij} \pi_{ij}}{2\sqrt{g}} \sigma   }
\hspace{2em} & & \\ 
{\displaystyle
+ 2 \int d^3y \frac{N \pi_{ij}}{\sqrt{g}} \frac{\delta}{\delta g_{ij}} 
        \int d^3z \sqrt{g} \tilde{\mathcal{V}}\,' \delta
+ 3 N \pi^{ij} \tilde{\mathcal{V}}\,'_{ij}  
    } 
& = & 0 \,.
\end{array}
\label{mueq}
\end{equation}

Equations (\ref{sigmaeq} - \ref{mueq}) form a system of equations for the Lagrange multipliers $\sigma$ and $\mu$. Indeed, since the potential depends on derivatives of $N$ and $g_{ij}$, the first two terms of both (\ref{sigmaeq}) and (\ref{mueq}) lead to differential operators on these multipliers. Thus, Eqs. (\ref{sigmaeq} - \ref{mueq}) are a coupled system of partial differential equations (PDEs) for $\sigma$ and $\mu$. Whenever this system can be solved for $\sigma$ and $\mu$, Dirac's algorithm for the preservation of constraints ends consistently with these equations (in the complete theory with $\lambda\neq 1/3$ Dirac's procedure ends with a PDE for $\sigma$ \cite{Donnelly:2011df,Bellorin:2011ff}).

Moreover, for the consistency of the whole Hamiltonian formulation it is of central importance that the Eqs.~(\ref{sigmaeq} - \ref{mueq}) can be solved for the multipliers $\sigma$ and $\mu$, without any further restriction on the canonical variables. Since the most relevant terms to determine the existence of solutions are the highest-derivative terms, it is illustrative to study the structure of the highest-order terms that arise in Eqs.~(\ref{sigmaeq} - \ref{mueq}) when the full (up to $z=3$) potential of the theory is considered. There is a big number of inequivalent $z=3$ operators that can be constructed with the curvature tensor and $a_i$; thus, at first sight it seems a very difficult task to elucidate the structure of the highest-order terms of Eqs.~(\ref{sigmaeq} - \ref{mueq}). However, a direct analysis on these equations may convince ourselves that some of these operators lead to the cube of the Laplacian, $\nabla^6$, acting on $\sigma$ or $\mu$ and the other ones lead to lower-order terms. The main point is that in Eqs.~(\ref{sigmaeq} - \ref{mueq}) the four terms
\begin{equation}
\begin{array}{c}
 {\displaystyle \int d^3y
  \,\sigma\, \frac{\delta}{\delta N} \int d^3z \sqrt{g} \tilde{\mathcal{V}} \delta \,,
\hspace{2em}
  \int d^3y \,\mu\, g_{ij} \frac{\delta} 
       {\delta g_{ij}} \int d^3z \sqrt{g} \tilde{\mathcal{V}} \delta \,, }
\\[2ex]
 {\displaystyle \int d^3y \,\mu\, g_{ij} \frac{\delta}{\delta g_{ij}} 
     \int d^3z \sqrt{g} \tilde{\mathcal{V}}\,' \delta \,,  
\hspace{2em}
 \int d^3y \,\sigma\, \frac{\delta}{\delta N} 
     \int d^3z \sqrt{g} \tilde{\mathcal{V}}\,' \delta }
\end{array}
\label{listterms}
\end{equation}
contain derivatives of the potential $\mathcal{V}$ of at least second order and there are $z=3$ operators that combine all their derivatives into a sixth-order derivative when they are functionally derived twice (or more). Let us consider, for example, the two operators $\mathcal{O}_1 \equiv (\nabla_i R_{jk})^2$ and $\mathcal{O}_2 \equiv \nabla_i a^i \nabla^2 \nabla_j a^j$. When $\mathcal{O}_1$ is inserted in the third term of (\ref{listterms}), the highest derivative that it gives rise on the Dirac delta is $\nabla^6$. After integration by parts, we get that $\mathcal{O}_1$ yields a term proportional to $\nabla^6 \mu$ in Eq.~(\ref{mueq}), whereas all its other contributions are of lower order. Similarly, $\mathcal{O}_2$ yields $\nabla^6$ acting on the Dirac delta when inserted in the first term of (\ref{listterms}), hence it leads to a term proportional to $\nabla^6 \sigma$ in Eq.~(\ref{sigmaeq}). The operators $\mathcal{O}_1$ and $\mathcal{O}_2$ must be included in the potential in their direct forms or in terms of their equivalent operators; that is, other operators that are obtained from them by integration by parts. In any case we obtain $\nabla^6$ as the highest-order operator acting on the Dirac delta once the functional derivatives in (\ref{listterms}) are performed. Other inequivalent operators yield the same result when inserted in some of the terms in (\ref{listterms}). We could anticipate this result by noting that there is no covariant differential operator of sixth order acting on the Dirac delta other than the cube of the Laplacian. Note, however, that there are $z=3$ terms that do not yield sixth-order derivatives acting on the delta when inserted in (\ref{listterms}), but lower-order derivatives. An example is $(a_i a^i)^3$. Another interesting example is the square Cotton, $C^{ij} C_{ij}$, which we shall consider explicitly in the next section.

In despite of the fact that some operators yield lower-order derivatives on $\sigma$ and $\mu$ in Eqs.~(\ref{sigmaeq} - \ref{mueq}), the ones that yield $\nabla^6$ \emph{must be included} in the potential and these are the dominant ones in Eqs.~(\ref{sigmaeq} - \ref{mueq}). After all the terms proportional to $\nabla^6 \sigma$ and $\nabla^6 \mu$ are collected, the last requisite is to impose that the matrix of their coefficients is positive definite, which is a condition in the space of coupling constants\footnote{Notice that there is no place for nonconstant coefficients in these terms. The potential only depends explicitly on $g_{ij}$ and $a_i$ and their derivatives; any coefficient that depends on them necessarily would increase the order, and we are considering the highest-order terms.}. Thus, we have that, when the full potential of the theory is considered, Eqs.~(\ref{sigmaeq} - \ref{mueq}) constitute a system of sixth-order, linear, elliptic PDEs for $\sigma$ and $\mu$ characterized by the highest-order terms $\nabla^6 \sigma$ and $\nabla^6 \mu$.

Having seen that the theory has a closed structure of constraints, we now evaluate the number of degrees of freedom. The theory has the first-class constraint $\mathcal{H}^i = 0$ and the second-class ones $\phi = \pi = \mathcal{H} = \mathcal{C} = 0$. They leave four independent degrees of freedom in the phase space, which correspond to the propagation of two even physical degrees of freedom. This is the same number of GR; \emph{there are not extra degrees of freedom in the complete, nonprojectable Ho\v{r}ava theory at the value $\lambda = 1/3$}. We may regard  the Hamiltonian constraint $\mathcal{H} = 0$ and the constraint $\mathcal{C}=0$, where $\mathcal{H}$ is given in (\ref{hamiltonianconstraintgeneral}) and $\mathcal{C}$ in (\ref{cconstraint}), as a coupled system of PDEs for the lapse function $N$ and one mode coming from the spatial metric $g_{ij}$.


\section{Soft breaking of the conformal symmetry}
\subsection{The full model}
In order to develop explicit computations, in this section we consider a model with $z=3$ and $z=1$ terms. This model will help us in clarifying the structure of the above equations. The quadratic ($z=1$) terms we consider are the most general ones; they are grouped in $\mathcal{V}^{(2)}$, which is given in (\ref{quadraticpotential}). The  importance of this quadratic potential lies on the fact that it is the leading one at the lowest energies, such that we build the lowest-order IR effective action with it and the kinetic term. This is the appropriated scenario to test whether GR is recovered at the IR. For the $z=3$ term we consider the square-Cotton term, which was the original $z=3$ term proposed by Ho\v{r}ava \cite{Horava:2009uw} and can be elegantly justified in the 3+1 action by the detailed balance principle.  

The potential is given by
\begin{equation}
 \mathcal{V} = - R - \alpha a_i a^i + w C_{ij} C^{ij} \,,
\label{model}
\end{equation}
where $w$ is a coupling constant and $C^{ij}$ is the Cotton tensor,
\begin{equation}
 C^{ij} = \frac{1}{\sqrt{g}} \varepsilon^{kl(i} \nabla_k R_l{}^{j)} \,.
\end{equation}
Computations with the $C^2$ term are facilitated by the fact that $\sqrt{g} C^{ij} C_{ij}$ transforms homogeneously (with weight $-3/2$) under conformal transformations of the metric, which are generated by $\pi$.

The modified potential (\ref{modifiedpotential}) and its derivative $\tilde{\mathcal{V}}\,'$ (\ref{vprima}) take the form (before imposing constraints)
\begin{eqnarray}
\tilde{\mathcal{V}} & = & 
- R + \alpha \left( 2 \nabla_i a^i + a_i a^i \right)  + w C_{ij} C^{ij} \,,
\\
\tilde{\mathcal{V}}\,' & = & 
- \frac{1}{2} N R + 2 N \nabla_i a^i + ( 2 - \alpha/2 ) N a_i a^i 
- \frac{3w}{2} N C_{ij} C^{ij} \,,
\end{eqnarray}
such that the Hamiltonian constraint (\ref{hamiltonianconstraintgeneral}) and the $\mathcal{C}$ constraint (\ref{cconstraint}) become
\begin{eqnarray}
\mathcal{H} &=& \frac{1}{\sqrt{g}} \pi^{ij} \pi_{ij} - \sqrt{g} R 
 + \alpha \sqrt{g} \left( 2 \nabla_i a^i + a_i a^i \right)
 + w \sqrt{g} C_{ij} C^{ij} = 0\,,
\label{hamiltonianconstraintmodel}
\\
\mathcal{C} &=&
  \frac{3N}{2\sqrt{g}} \pi^{ij} \pi_{ij}
+ \frac{1}{2} \sqrt{g} N R 
- \sqrt{g} N \left( 2  \nabla_i a^i
+ ( 2 - \alpha/2 ) a_i a^i \right)
+ \frac{3w}{2} \sqrt{g} N C_{ij} C^{ij} = 0\,. \nonumber \\ 
\label{cmodel}
\end{eqnarray}
Finally, we evaluate the equations (\ref{sigmaeq} - \ref{mueq}) for the model (\ref{model}). As we anticipate in the previous section, these equations lead to a coupled system of PDEs for $\sigma$ and $\mu$,
\begin{eqnarray}
  \beta \left( 2 \nabla^2 \sigma + N a^i \partial_i \mu \right)
- 2 g^{-1} \pi^{ij} \pi_{ij} \sigma
+ \left( \beta \nabla^2 N + 3 g^{-1} N \pi^{ij} \pi_{ij} \right) \mu
+ ( 3 \mu N - 2 \sigma ) w C^{ij} C_{ij} = &&
\nonumber \\
 - \frac{4 N}{\sqrt{g}} \pi^{ij} ( N R_{ij} -\nabla_i \nabla_j N  
   + \alpha N a_i a_j ) 
 + \frac{4\beta}{\sqrt{g}} \partial_i ( N \partial_j N \pi^{ij} ) 
&& \nonumber \\
 - \frac{8 w N}{\sqrt{g}} C_{ij} \mathbb{O}^{ijkl} ( N \pi_{kl} )
 + \frac{8 w N}{\sqrt{g}} \pi_{kl} \bar{\mathbb{O}}^{ijkl} ( N C_{ij} )  \,, && \nonumber \\ 
\label{sigmaeqmodel}
\\
  \nabla^2 \mu 
- \frac{\alpha}{N} a^i \partial_i \sigma
- \frac{1}{4} \left(
   R + \alpha a_i a^i + (3/\gamma) g^{-1} \pi^{ij} \pi_{ij} \right) \mu
+ \frac{\alpha}{N} a_i a^i \sigma  
- \frac{3 w}{4\gamma} C_{ij} C^{ij} \mu
= \,\,\, \hspace{2em} &&
\nonumber \\
 \frac{2\alpha}{\beta \sqrt{g}} \pi^{ij} \left( N R_{ij} 
     - \nabla_i \nabla_j N + \alpha N a_i a_j \right) 
- \frac{2 w}{\gamma\sqrt{g}} C_{ij} \mathbb{O}^{ijkl} ( N \pi_{kl} )
+ \frac{2 w}{\gamma\sqrt{g}} \pi_{kl} \bar{\mathbb{O}}^{ijkl} ( N C_{ij})
\,, && \nonumber \\
\label{mueqmodel}
\end{eqnarray}
where $\mathbb{O}^{ijkl}$ and $\bar{\mathbb{O}}^{ijkl}$ are the differential operators
\begin{equation}
 \begin{array}{rcl}
  \mathbb{O}^{ijkl} & \equiv & 
  {\displaystyle\frac{1}{2\sqrt{g}} \left(
     \varepsilon^{imk} \nabla_m \nabla^l \nabla^j
   + g^{jl} \varepsilon^{imn} \nabla_m \nabla^k \nabla_n
   - g^{jl} \varepsilon^{imk} \nabla_m \nabla^n \nabla_n
   - g^{kl} \varepsilon^{imn} \nabla_m \nabla^j \nabla_n
\right)} \\[2ex] &  &
   + {\displaystyle\frac{1}{2\sqrt{g}}\left(
       g^{jk} \varepsilon^{iln} R_n{}^m \nabla_m
     - g^{jk} \varepsilon^{imn} R_n{}^l \nabla_m
     - g^{jm} \varepsilon^{ikn} R_n{}^l \nabla_m
\right)} \,,
\\[2ex]
\bar{\mathbb{O}}^{ijkl} & \equiv & 
  {\displaystyle\frac{1}{2\sqrt{g}} \left(
   - \varepsilon^{imk} \nabla^j \nabla^l \nabla_m 
   - g^{jl} \varepsilon^{imn} \nabla_n \nabla^k \nabla_m 
   + g^{jl} \varepsilon^{imk} \nabla_n \nabla^n \nabla_m 
   + g^{kl} \varepsilon^{imn} \nabla_n \nabla^j \nabla_m
\right)} \\[2ex] &  & 
  + {\displaystyle\frac{1}{2\sqrt{g}} \nabla_m \left[\left(
   - g^{jk} \varepsilon^{iln} R_n{}^m
   + g^{jk} \varepsilon^{imn} R_n{}^l
   + g^{jm} \varepsilon^{ikn} R_n{}^l
   \right)\cdot\right]} \,,
 \end{array}
\end{equation}
and
\begin{equation}
\beta \equiv ( 1 - \alpha/2 ) \,,
\hspace{2em}
\gamma \equiv \left( \frac{ 1 - \alpha/2 }{ 1 + 3\alpha/2} \right) \,.
\end{equation}
We have made some massaging on Eqs.~(\ref{sigmaeqmodel} - \ref{mueqmodel}) to bring them to their final form, in particular by using the constraints (\ref{hamiltonianconstraintmodel} - \ref{cmodel}). Equations (\ref{sigmaeqmodel} - \ref{mueqmodel}) constitute a system of linear, elliptic, PDEs for the Lagrange multipliers $\sigma$ and $\mu$. Notice that these equations are of second order although the model is $z=3$. This is because the $\sqrt{g} N C_{ij} C^{ij}$ term is covariant under conformal transformations generated by $N\phi$ and $\pi$. Thus, the Poisson brackets between $\sqrt{g} N C_{ij} C^{ij}$ and $\sigma\phi$, $\mu \pi$ generate terms proportional to the $C^2$ term, $\sigma$ and $\mu$ playing the role of conformal factors. Hence no derivatives of $\sigma$ or $\mu$ are generated by the $C^2$ term.  Equations (\ref{sigmaeqmodel} - \ref{mueqmodel}) can be solved for $\sigma$ and $\mu$ on general grounds, hence we conclude that Dirac's algorithm for the preservation of constraints ends consistently with these equations.

As we pointed out in the previous section, constraints (\ref{hamiltonianconstraintmodel} - \ref{cmodel}) can be regarded as a system of PDEs for $N$ and one mode coming from $g_{ij}$. We would like to emphasize this point for $N$. Constraints (\ref{hamiltonianconstraintmodel} - \ref{cmodel}) can be combined to obtain the equation
\begin{equation}
 \beta \nabla^2 N = 
 N \left( \frac{1}{g} \pi^{ij} \pi_{ij} + w C_{ij} C^{ij} \right)
\label{neqmodel}
\end{equation}
The structure of this equation falls on the same class of equations we studied in Ref.~\cite{Bellorin:2012di}. There we showed that, if the term without derivatives of $N$ satisfies a positiveness condition, then the solution for $N$, in the sense of distributions, exists and is unique. For the Eq.~(\ref{neqmodel}), the nonderivative term is definite positive by requiring only $w\geq 0$. Assuming also $\alpha < 2$ ($\beta > 0$), then the results of Ref.~\cite{Bellorin:2012di} imply the existence and uniqueness of $N$, and moreover, it is guarantied that $N\geq 0$ over all the spatial submanifold.

 By evaluating conditions (\ref{inequalities}) (or directly from (\ref{hamiltonianconstraintmodel} - \ref{cmodel})), we get, imposing $w \geq 0$,
\begin{eqnarray}
R - (1+3\alpha/2) \nabla_i a^i - (1+\alpha/2) a_i a^i & = & 0 \,,
\\
R - 2 \alpha \nabla_i a^i - \alpha a_i a^i & \geq & 0 \,.
\end{eqnarray}
It is straightforward to deduce from these relations the following inequalities
\begin{equation}
\beta \nabla^2 N \geq 0 \,,
\hspace{2em}
\gamma \left( R + \alpha a_i a^i \right) \geq 0 \,.
\label{positive}
\end{equation}

On the basis of the inequalities (\ref{positive}), we may give a result about the positiveness of the energy of the model (\ref{model}). If the coupling constant $\alpha$ is restricted to the set
\begin{equation}
 -\frac{2}{3} < \alpha \leq 0 \,,
\label{setalpha}
\end{equation}
then the inequalities (\ref{positive}) yield $\nabla^2 N \geq 0$ and $R \geq 0$. The former implies that the flux of $N$ at infinity is nonnegative. The positive energy theorem of GR \cite{Schoen,Witten} establishes that an everywhere positive $R$ gives a positive ADM energy. Therefore, the energy of the model, given in (\ref{e}),
\begin{equation}
 E = E_{\mbox{\tiny ADM}} - 2 \alpha \Phi_N \,,
\end{equation} 
is nonnegative when $\alpha$ is restricted to the range (\ref{setalpha}).

\subsection{The IR effective action}
Now we move to the $w\rightarrow 0$ limit to extract the properties of the IR effective action of the complete Ho\v{r}ava theory at the $\lambda =1/3$ value, which has the potential $\mathcal{V}^{(2)} = - R - \alpha a_i a^i$. This effective theory has a consistent structure of constraints by itself, which can be seen by taking the $w\rightarrow 0$ limit on the constraints/equations for multipliers of the above model. We obtain the $\mathcal{H}$ and $\mathcal{C}$ constraints
\begin{eqnarray}
\mathcal{H} &=& \frac{1}{\sqrt{g}} \pi^{ij} \pi_{ij} - \sqrt{g} R 
 + \alpha \sqrt{g} \left( 2 \nabla_i a^i + a_i a^i \right)
 = 0\,,
\label{hamiltonianconstraint}
\\
\mathcal{C} &=&
  \frac{3N}{2\sqrt{g}} \pi^{ij} \pi_{ij}
+ \frac{1}{2} \sqrt{g} N R 
- \sqrt{g} N \left( 2  \nabla_i a^i
+ ( 2 - \alpha/2 ) a_i a^i \right)
 = 0\,,
\label{c}
\end{eqnarray}
and we notice that the Eqs.~(\ref{sigmaeqmodel} - \ref{mueqmodel}) for $\sigma$ and $\mu$ maintain their structure of second order, linear, elliptic PDEs. By evaluating the Eqs.~(\ref{inequalities}), we obtain again the Eqs.~(\ref{positive}) in the same form, hence the result about the positiveness of the energy of the IR effective action holds.

Since the number of propagating degrees of freedom is the same of GR, it is interesting to elucidate whether GR can be recovered at the lowest energies from this model of Ho\v{r}ava theory. The best way to analyze this point is the perturbative analysis. We perform perturbations of the Minkowski background in the way
\begin{equation}
 g_{ij} = \delta_{ij} + h_{ij} \,,
\hspace{2em}
 \pi^{ij} = p_{ij} \,,
\hspace{2em}
 N = 1 + n \,.
\end{equation}
We decompose the perturbative metric into transverse and longitudinal parts,
\begin{equation}
 h_{ij} = 
 h_{ij}^{TT} 
 + {\textstyle\frac{1}{2}}[ \delta_{ij} 
    - \partial_{ij} (\partial_{kk})^{-1}] h^T
 + \partial_i h^L_j + \partial_j h^L_i 
 + \partial_{ij} (\partial_{kk})^{-1} h^L \,,
\label{decomposition}
\end{equation}
where the boundary conditions are used in the definition of $(\partial_{kk})^{-1}$. In the above we use the shorthand $\partial_{ij} = \partial_i \partial_j$ and so on. We also decompose $p_{ij}$ in the same way of (\ref{decomposition}). In addition, we impose the gauge in which all the longitudinal sector of the metric is eliminated,
\begin{equation}
 \partial_i h_{ij} = 0 \,.
\label{gauge}
\end{equation}

We start the perturbations by analyzing the constraints of the theory at linear order. The linear-order momentum constraint $\mathcal{H}^i = 0$ (\ref{momentumconstraint}) yields
\begin{equation}
 \partial_i p_{ij} = 0 \,.
\end{equation}
Hence the longitudinal sector of $p_{ij}$ is also eliminated. Constraint (\ref{picero}) yields $p^T = 0$. This leaves us with only the sector $p_{ij}^{TT}$ activated. The perturbative Hamiltonian constraint $\mathcal{H}= 0$ (\ref{hamiltonianconstraint}) and the constraint $\mathcal{C} = 0$ (\ref{c}) at linear order yield, respectively,
\begin{eqnarray}
 \partial_{ii} h^T + 2\alpha \partial_{ii} n & = & 0 \,,
\\
 \partial_{ii} h^T + 4 \partial_{ii} n & = & 0 \,.
\end{eqnarray}
With the prescribed boundary conditions, and assuming $\alpha \neq 2$, there are not solutions to these equations other than 
\begin{equation}
 h^T = n = 0 \,.
\end{equation}
Thus, we see that at linear order the transverse scalar mode of the metric and the lapse function are switched off completely. Similarly, the linearized equations (\ref{sigmaeq} - \ref{mueq}) yield $\sigma = \mu = 0$.

After all the linear-order constraints are imposed, the unconstrained second-order Hamiltonian (\ref{hamiltonianfinal}) takes the form
\begin{equation}
 H = \int d^3 x \left(
  p^{TT}_{ij} p^{TT}_{ij}
 + \frac{1}{4} \partial_i h^{TT}_{jk} \partial_i h^{TT}_{jk} 
\right) \,,
\end{equation}
which is exactly the Hamiltonian of linearized GR. Thus, we see that the dynamics of linearized GR is smoothly and exactly recovered from the complete, nonprojectable Ho\v{r}ava theory at the value $\lambda =1/3$. There is not any discontinuity in the degrees of freedom since the nonperturbative theory has also two degrees of freedom.


\section{Conclusions}
We introduce a theory of gravitation based on the principle of having the foliation-preserving diffeomorphisms as gauge symmetry. The theory we consider can be obtained from the Ho\v{r}ava theory with the terms of Blas, Pujol\`as, and Sibiryakov by setting the coupling constant $\lambda$ equal to $1/3$. The terms of Blas, Pujol\`as, and Sibiryakov must be included to obtain a renormalizable theory since they are compatible with the gauge symmetry. Unlike other models of Ho\v{r}ava theory, in our model the value $\lambda=1/3$ is protected under quantum corrections because of the presence of a constraint in the theory. This implies that $\lambda$ is not actually a running coupling constant.

Our Hamiltonian analysis shows that the theory has a consistent and closed structure of constraints. It also provides elliptic equations for the elimination of the Lagrange multipliers associated to the primary second-class constraints. As a remarkable feature of the theory, we found that it has two second-class constraints that are absent in others models based on Ho\v{r}ava theory. One of them is $\pi = 0$, which is a primary constraint of the theory. These constraints get rid of the extra mode that arises in $\lambda\neq 1/3$ models of Ho\v{r}ava theory (with or without the terms of Blas, Pujol\`as, and Sibiryakov). We would like to stress that these constraints are always present for any choice of the potential, since they are a consequence of the universal form of the kinetic term of the Ho\v{r}ava theory and the value $\lambda=1/3$. As a result, we get that the theory has two physical degrees of freedom, as general relativity has.

We consider a concrete potential with a $z=3$ term, a square-Cotton term, and all the $z=1$ terms compatible with the gauge symmetry. The square-Cotton term can be obtained from the principle of detailed balance \cite{Horava:2009uw}. For this model, we additionally show that its energy is positive definite.

Another outstanding feature of the theory is that the linear-order perturbative version, around Minkowski space-time, of the IR effective action is physically equivalent to linearized general relativity. We obtain the IR effective action as the $z=1$ truncation of the concrete model, since it includes all the second-order terms (including the Blas, Pujol\`as, and Sibiryakov term). Thus, we have obtained a clear and consistent way to obtain general relativity in the low energy limit of Ho\v{r}ava theory.

We think that our model is a good candidate for a renormalizable theory of quantum gravity. Since the quantization must be performed on the constrained submanifold, the second-class constraint $\pi=0$ must be preserved and this avoids $\lambda$ to move from the value $\lambda=1/3$. However, one should be careful with the treatment of the second-class constraints when performing the quantization of the theory.


\section*{Acknowledgments}
A. R. and A. S. are partially supported by Project Fondecyt
1121103, Chile.


\end{document}